\documentclass[aps,prb,twocolumn,showpacs,superscriptaddress]{revtex4}

\usepackage{graphicx}

\begin{document}

\title{Two spatially separated phases in semiconducting Rb$_{0.8}$Fe$_{1.5}$S$_2$}
\author{Meng Wang}
\email{wangm@berkeley.edu}
\affiliation{Department of Physics, University of California, Berkeley, California 94720, USA }
\author{Wei Tian}
\affiliation{Quantum Condensed Matter Division, Oak Ridge National Laboratory, Oak
Ridge, Tennessee 37831-6393, USA}
\author{P. Valdivia}
\affiliation{Department of Physics, University of California, Berkeley, California 94720, USA }
\author{Songxue Chi}
\affiliation{Quantum Condensed Matter Division, Oak Ridge National Laboratory, Oak
Ridge, Tennessee 37831-6393, USA}
\author{E. Bourret-Courchesne}
\affiliation{Materials Science Division, Lawrence Berkeley National Laboratory, Berkeley, California 94720, USA }
\author{Pengcheng Dai}
\affiliation{Department of Physics and Astronomy, Rice University, Houston, Texas 77005, USA}
\affiliation{Institute of Physics, Chinese Academy of Sciences, P. O. Box 603, Beijing
100190, China }
\author{R. J. Birgeneau}
\affiliation{Department of Physics, University of California, Berkeley, California 94720, USA }
\affiliation{Materials Science Division, Lawrence Berkeley National Laboratory, Berkeley, California 94720, USA }
\affiliation{Department of Materials Science and Engineering, University of California, Berkeley, California 94720, USA }

\begin{abstract}
We report neutron scattering and transport measurements on semiconducting Rb$_{0.8}$Fe$_{1.5}$S$_2$, a compound isostructural and isoelectronic to the well-studied $A_{0.8}$Fe$_{y}$Se$_2 (A=$ K, Rb, Cs, Tl/K) superconducting systems.  Both resistivity and DC susceptibility measurements reveal a magnetic phase transition at $T=275$ K. Neutron diffraction studies show that the 275 K transition originates from a phase with rhombic iron vacancy order which exhibits an in-plane stripe antiferromagnetic ordering below 275 K. In addition, interdigitated mesoscopically with the rhombic phase is an ubiquitous phase with $\sqrt{5}\times\sqrt{5}$ iron vacancy order. This phase has a magnetic transition at $T_N=425$ K and an iron vacancy order-disorder transition at $T_{S}=600$ K. These two different structural phases are closely similar to those observed in the isomorphous Se materials. Based on the close similarities of the in-plane antiferromagnetic structures, moments sizes, and ordering temperatures in semiconducting Rb$_{0.8}$Fe$_{1.5}$S$_2$ and K$_{0.81}$Fe$_{1.58}$Se$_2$, we argue that the in-plane antiferromagnetic order arises from strong coupling between local moments. Superconductivity, previously observed in the $A_{0.8}$Fe$_{y}$Se$_{2-z}$S$_z$ system, is absent in Rb$_{0.8}$Fe$_{1.5}$S$_2$, which has a semiconducting ground state. The implied relationship between stripe/block antiferromagnetism and superconductivity in these materials as well as a strategy for further investigation is discussed in this paper. 

\end{abstract}

\maketitle




\section{Introduction}
The $A_{0.8}$Fe$_{1.6\pm\delta}$Se$_2$ ($A=$K, Rb, Cs, Tl/K) materials, the so-called `245' systems were discovered at the end of 2010, and have since generated a great deal of interest, in large part because of their unique properties: iron vacancy order, block antiferromagnetism (AF) with large 3.3 $\mu_B$ moments aligned along the $c-$axis, and the existence of superconductivity for appropriate chemical compositions\cite{jgguo, krzton, mhfang, afwang, wbao1, fye}. In the Fe pnictide systems, the parent compounds of the superconductors exhibit a collinear antiferromagnetic structure with small ordered moments, typically less than 1 $\mu_B$\cite{cruz, huang, shiliang, lynn}. Superconductivity arises upon electron or hole doping of the parent compounds, which concomitantly suppresses the AF order. Spin fluctuations associated with the AF order, which exist throughout the superconducting (SC) dome, are thought to play a crucial role in the mechanism of superconductivity\cite{johnston, dai, meng}. In the standard interpretation, nesting between the hole and electron Fermi surfaces gives rise to spin density wave (SDW) order. In addition, the ubiquitous occurrence of a neutron `spin-resonance' at the SDW wave vector in superconducting iron pnictide compounds has been suggested to correlate with `$s\pm$' pairing symmetry\cite{jdong, mazin, maier}.

A spin resonance mode was also found in the `245' system, but at a wave vector different from those of both the block and stripe AF orders\cite{park,gfriemel}. Importantly, unlike the Fe pnictides, a weak electron-like Fermi pocket and hole-like bands below the Fermi surface are found in place of hole Fermi surfaces around the $\Gamma$ point in the $A_{0.8}$Fe$_{1.6}$Se$_{2}$ system\cite{yzhang, xpwang, xjzhou}. The occurrence of superconductivity with $T_c$ up to 32 K in the `245' system in the absence of electron-hole nesting presents a significant challenge to current theories of these phenomena\cite{dagoto}. 

There is extensive empirical evidence that the SC phase occurs mesoscopically separated from the block AF insulator\cite{ricc1,ricc2,fchen, xue, wli,jqli,meng2}. The block AF phase exists throughout the two dimensional phase diagrams of $A_{x}$Fe$_y$Se$_2$ over wide variations in the alkali metal ($0.77 \le x\le 0.98$) and iron contents ($1.48 \le y \le 1.65$), with little change of $T_N$\cite{wbao3}. We emphasize that $(x, y)$ are for the sample as a whole, not the two separate constituent phases in most studies. The reports focused on the composition of the superconducting phase remain conflicting\cite{wli, lazarevic1, jzhao, scott}. Thus, the nature of the real superconducting phase and its parent compound are still under debate\cite{fchen, xue, wli, lazarevic1,wen, scott, jzhao}. Both theory and photoemission experiment proposed an insulating or semiconducting phase as a candidate for the parent compound of the superconducting phase in (K, Tl)$_x$Fe$_y$Se$_2$\cite{si, fchen}. Importantly, the same stripe AF structure with in-plane ordered moments that occurs in the parent compounds of pnictide superconductors was observed in semiconducting K$_{0.81}$Fe$_{1.58}$Se$_2$ by neutron diffraction\cite{jzhao}. If the stripe phase with in-plane AF order is, in fact, the parent compound of the superconducting phase in the `245' system, then the SC in this system may have the same underlying mechanism as that in the other iron based superconductors, in spite of the absence of electron-hole nesting and different neutron spin resonance wave vectors\cite{jdong,mazin,maier,park,gfriemel,yzhang,xpwang,xjzhou}.  Therefore, determining the origin of the in-plane AF order in the semiconducting phase and its relationship with superconductivity is crucial to understanding the mechanism of superconductivity in the $A_x$Fe$_y$Se$_2$ system.

The low-temperature electrical resistivity of  the `245' system can be changed from insulating to semiconducting or superconducting by controlling the iron content as in $A_{0.8}$Fe$_y$Se$_2$, generally in concert with the alkali concentration $A$, or by substitution of sulfur on the selenium sites as in $A_{0.8}$Fe$_y$Se$_{2-z}$S$_z$\cite{gengfu, jgguo2, kefeng, hechang, caiyao, jzhao}. In studies to-date, changing the iron content of the pure Se two-phase material results in the sudden disappearance of the superconductivity, while sulfur substitution for selenium appears to suppress superconductivity gradually, resulting in a semiconducting ground state\cite{hechang}. Accordingly, semiconducting $A_{0.8}$Fe$_y$S$_2$ may also be viewed as the parent compound of the $A_{0.8}$Fe$_y$Se$_{2-z}$S$_z$ superconductors, although the magnetic phase diagram has not yet been determined for high sulfur substitution. Both high temperature transport and Raman scattering measurements indicate that the block AF phase also exists in the $A_{x}$Fe$_y$S$_2$ system\cite{jjying, lazarevic}. Thus, it is important to investigate whether or not the in-plane AF order occurs in $A_{0.8}$Fe$_y$S$_2$ and, if so, to determine its relationships with superconductivity in the S-substituted $A_{0.8}$Fe$_y$Se$_{2-z}$S$_{z}$. 

\begin{figure}[t]
\includegraphics[scale=0.4]{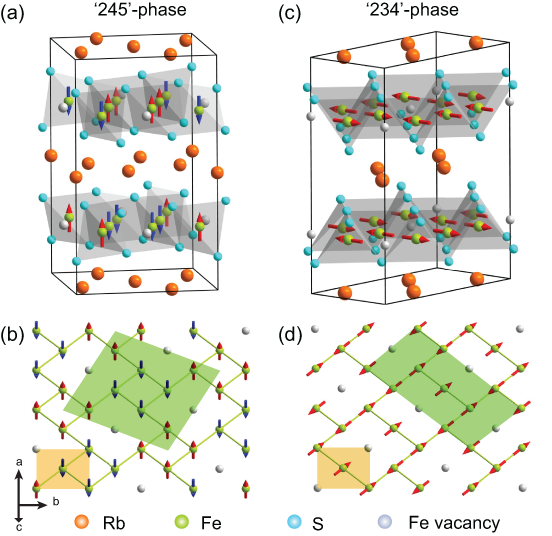}
\caption{ (color online). (a)  Three dimensional structure of the block AF phase in Rb$_{0.8}$Fe$_{1.5}$S$_{2}$. (b) The Fe plane of block AF phase with $\sqrt{5}\times\sqrt{5}$ iron order, where a magnetic unit cell with lattice parameter $a_s=\sqrt{5}\times3.765$ \AA\ has been marked as green. The tetragonal lattice cell used throughout this paper is shaded light orange. The red and blue arrows represent the out-of-plane spin directions up and down. The orange, lime, turquoise and light gray balls are Rb, Fe, S atoms and Fe vacancies, respectively. (c) A three-dimensional magnetic unit cell of the in-plane AF order and (d) Fe-plane with the rhombic iron vacancy order. A magnetic unit cell is shaded green. The diagonal Fe bonds are 3.889 \AA\ at 180 K.  }
\label{fig1}
\end{figure}

In this paper, we present transport and elastic neutron scattering measurements on single crystals of semiconducting Rb$_{0.8}$Fe$_{1.5}$S$_2$. Two magnetic phases are found in this material with the next nearest (NN) Fe neighbor bond distances at 180 K 3.765 \AA\ and 3.889 \AA\ for the two phases respectively. The first phase, the `245' phase, which has the more compact in-plane lattice constants, has the $\sqrt{5}\times\sqrt{5}$ iron vacancy order and block AF order as in the $A_{0.8}$Fe$_{1.6\pm\delta}$Se$_2$ system\cite{wbao3}.  The N$\acute{\mathrm{e}}$el temperature of the block AF order is 425 K; this is reduced significantly compared with $\sim$560 K in $A_{0.8}$Fe$_y$Se$_2$, and also well separated from the $\sqrt{5}\times\sqrt{5}$ iron vacancy ordering temperature of 600 K in Rb$_{0.8}$Fe$_{1.5}$S$_2$\cite{wbao1}. Schematics of the three-dimensional  structure together with that of the iron plane with ordered moments and iron vacancies are shown in Fig.~\ref{fig1} (a, b). The second phase has rhombic iron vacancy order with in-plane stripe AF order below 275 K (Fig.~\ref{fig1} (c, d))\cite{sabrowsky}. We named it as the `234' phase (this assumes an ideal stoichiometry RbFe$_{1.5}$S$_2$) in spite of the possible deviation of Rb in the discussion below. The estimated  in-plane magnetic moment size of  $2.8\pm0.5 \mu_B$ and the  N$\acute{\mathrm{e}}$el temperature of 275 K for the stripe AF order in semiconducting Rb$_{0.8}$Fe$_{1.5}$S$_2$ are surprisingly close to the 2.88 $\mu_B$ moments and $T_N=280$ K of the stripe AF order in semiconducting K$_{0.81}$Fe$_{1.58}$Se$_2$\cite{jzhao}. These results suggest that strong coupling of local moments plays the dominant role in the formation of in-plane AF order in semiconducting $A_{0.8}$Fe$_yX_2$ ($X=$Se, S).

\section{Experimental Details}

Our experiments were carried out on the HB-1A triple-axis spectrometer and HB-2C wide-angle neutron diffractometer (WAND) at the High-Flux Isotope Reactor, Oak Ridge National laboratory. The triple-axis experiment employed two pyrolytic graphite (PG) filters before the sample to reduce $\lambda/2$ contamination and horizontal collimation 40$^{\prime}$-40$^{\prime}$-$S$-40$^{\prime}$-80$^{\prime}$ with a fixed incident beam energy of $E_i=14.64$ meV. A single piece of crystal weighing 220 mg with a mosaic of 1.5$^{\circ}$ was loaded into a closed cycle refrigerator (CCR) which covers the temperature range from 30 K to 750 K. The sample was aligned in the $[H, H, L]$ zone and the $[H, 3H, L]$ zone in tetragonal notation with lattice parameters $a=b=3.889$ \AA, $c=13.889$ \AA\ for the `234' phase and  $a=b=3.765$ \AA, $c=13.889$ \AA\ for the `245' phase optimized at 180 K. The momentum transfer $(q_x, q_y, q_z)$ is defined as $(2\pi H/a, 2\pi K/b, 2\pi L/c)$ in \AA$^{-1}$. ($H, K, L$) are the Miller indices in reciprocal lattice units (r.l.u). We labeled the wave vectors as $Q=[H, K, L]$ r.l.u throughout this paper. By employing the degrees of freedom of the upper and lower goniometers of HB-1A, we were able to probe wave vectors in the $[H, K, L]$ plane near $[H, H, L]$. A Ge(115) monochromator was used to produce a neutron beam with $\lambda=1.482$ \AA\ in the experiment at WAND. The one-dimensional $^3$He detectors with 624 anodes can cover a wide range in reciprocal space by rotating the sample. A standard CCR was used to cover the temperature range between 4 K and 300 K.

The Rb$_{0.8}$Fe$_{1.5}$S$_{2}$ single crystals were grown by the Bridgman method with a one-step reacting. Stoichiometric amounts of high purity of a Rb ingot, Fe powder, and pieces of S were loaded in an alumina crucible in an argon gas filled glovebox; then the alumina crucible was sealed in a quartz tube under vacuum. The quartz tube was loaded into a box furnace and kept at 200 $^{\circ}$C for 24 hours; then warmed up to 500 $^{\circ}$C and held for 20 hours; heated slowly to 1050 $^{\circ}$C for melting 5 hours; cooled down to 750 $^{\circ}$C at a rate of 4 $^{\circ}$C/hour; and finally cooled to room temperature. We obtained single crystals with dimensions up to $5\times5\times4$ mm$^{3}$.

\section{Results}

\begin{figure}[t]
\includegraphics[scale=0.6]{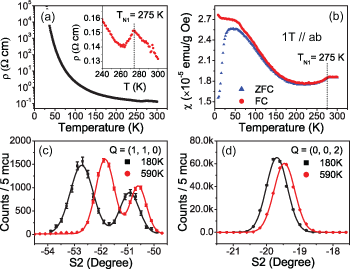}
\caption{ (color online). (a) In-plane resistivity measurement of semiconducting Rb$_{0.8}$Fe$_{1.5}$S$_2$.  (b) 1 Tesla Zero field cooled (ZFC) and field cooled (FC) dc magnetic susceptibility measurements with $H\parallel ab$-plane. A spin-glass like behavior appears below 55 K.  A kink corresponding to the stripe AF order transition is observed in both resistivity and susceptibility measurements at $T=$275 K. (c) The well separated peaks of $S2$ scans at nuclear reflection wave vector $Q=(1, 1, 0)$  demonstrate two different sets of in-plane lattice constants at 180 K and 590 K. (d)  $\theta$-$2\theta$ scans at $Q=(0, 0, 2)$ show a single lattice constant $c$ within the instrument resolution. The error bars are one standard deviation and solid lines are fits to Gaussian function throughout this paper. }
\label{fig2}
\end{figure}

We characterized the transport properties of several Rb$_{0.8}$Fe$_{1.5}$S$_{2}$ single crystals with a Quantum Design Physical Property Measurement System (PPMS). The results were very consistent among the different samples measured and indicated consistent phases. The in-plane resistivity shown in Fig.~\ref{fig2} (a) on a logarithm scale represents clear semiconducting behavior. This semiconducting characteristic is quite similar to that of the potassium compound with equivalent composition, K$_{0.8}$Fe$_{1.5}$S$_2$\cite{caiyao}. These results reveal, as expected, that Rb$_{0.8}$Fe$_{y}$S$_{2}$ and K$_{0.8}$Fe$_{y}$S$_{2}$ have similar transport characteristics. The enlarged resistivity from 240 K to 300 K in the inset of Fig.~\ref{fig2} (a) implies a phase transition at 275 K. This transition temperature corresponds to the onset of the in-plane stripe AF order observed by neutron diffraction, which is discussed in more detail below. The kink at 275 K corresponding to the stripe AF transition can also be seen in the susceptibility measurement. The difference between the zero field cooled (ZFC) and field cooled (FC) susceptibilities in Fig.~\ref{fig2} (b) suggests the possibility of a spin-glass phase coincident with the long range AF order in Rb$_{0.8}$Fe$_{1.5}$S$_{2}$, similar to that which has been proposed for K$_{0.88}$Fe$_{1.63}$S$_2$\cite{hechang2}. 

The lattice constants can be optimized by carrying out longitudinal scans: scanning the angle of the incident beam and exit beam $S2$ ($2\theta$), and rotating the sample angle ($\theta$) by half of the step. The two well separated peaks of the longitudinal scan at $Q=(1, 1, 0)$ are strong evidence for two structural phases existing in this sample[Fig.~\ref{fig2} (c)]. As estimated from the integrated peak intensities at 180 K, the `245' phase with $a=b=3.765$ \AA\ (peak centered at $-$52.70) has $\sim$65\% volume fraction, and the `234' phase  with $a=b=3.889$ \AA\  (peak centered at $-$50.90)  has $\sim$35\% volume fraction. We observed that the volume fractions of the two phases varied among our different samples\cite{discuss}. The transition temperatures of each phase are consistent. The two phases have the same lattice constant $c=13.889$ \AA\, based on the $\theta$-$2\theta$ scans at $Q=(0, 0, 2)$ at 180 K as shown in Fig.~\ref{fig2} (d). The peaks shift slightly due to the change of lattice constants at 590 K, but the peaks at $Q=(1, 1, 0)$ are still clearly distinguishable at 590 K\cite{discuss}. This is in marked contrast with the behavior in phase-separated superconducting K$_{0.8}$Fe$_{1.6}$Se$_2$, where the non-magnetic phase with the more compact in-plane lattice constant merges together with the block AF phase at temperatures above the iron vacancy order-disorder transition at 520 K\cite{ricc1, scott}.

\begin{figure}[t]
\includegraphics[scale=0.6]{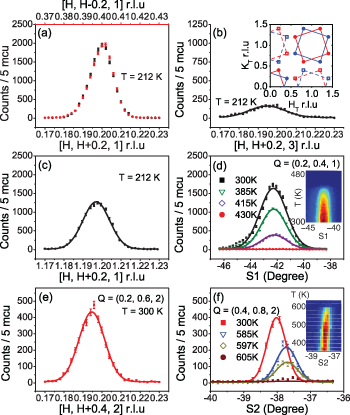}
\caption{(color online). Magnetic and nuclear reflection peaks associated with the block AF phase and $\sqrt{5}\times\sqrt{5}$ iron vacancy order. Scans through magnetic peaks at (a) $Q=(0.2, 0.4, 1)$, (b) $Q=(0.2, 0.4, 3)$ and (c) $Q=(1.2, 1.4, 1)$ along $[H, H+0.2]$ direction at $T=212$ K. The red solid points in (a) through $Q=(0.4, 0.2, 1)$ are from the other chirality. (d) Rocking curve scans through magnetic peak $Q=(0.2, 0.4, 1)$ at selected temperatures of $T=$ 300, 385, 415 and 430 K. (e) Scan through the nuclear peak at $Q=(0.2, 0.6, 2)$ along $[H, H+0.4]$ direction associated with the $\sqrt{5}\times\sqrt{5}$ iron vacancy order at $T=300$ K. (f) $\theta$-$2\theta$ scans through nuclear peak at $Q=(0.4, 0.8, 2)$ at $T=$ 300, 585, 597 and 605 K.  The inset in (b) is the expected magnetic Bragg peaks (solid circles, $L=$ odd) and nuclear Bragg peaks (empty squares, $L=$ even)  in tetragonal reciprocal space. The red and blue lines indicate two different chiralities. The insets in (d) and (f) are color maps of temperature dependence of reflection peaks corresponding to (d) and (f). }
\label{fig3}
\end{figure}

We first discuss the `245' phase, which has the block AF order and $\sqrt{5}\times\sqrt{5}$ iron vacancy order in Fig.~\ref{fig1} (a, b). The block AF order generates magnetic peaks at the wave vectors shown as the solid circles in the scattering planes of $[H, K, L=odd]$ in the inset figure of Fig.~\ref{fig3} (b). The $\sqrt{5}\times\sqrt{5}$ iron vacancy order produces nuclear peaks at the positions of the blank squares in the $[H, K, L=even]$ planes. The wave vectors connected by red and blue lines in the inset of Fig.~\ref{fig3} (b) originate from the left and right chiralities, respectively. The details of the diffraction have been described elsewhere\cite{meng2}. By comparing the peak centers under the two sets of lattice constants, the set with $a=b=3.765$ \AA\ was determined to correspond to the block AF phase. Fig.~\ref{fig3} (a-c) represent scans at the magnetic wave vectors of $Q=(0.2, 0.4, 1)$, $Q=(0.2, 0.4, 3)$ and $Q=(1.2, 1.4, 1)$ at 212 K. The dramatic decrease of the magnetic peak intensity at $L=3$ compared with that at $L=1$ is consistent with $c$-axis-aligned moments together with the Fe$^{2+}$ magnetic form factor.  Fig.~\ref{fig3} (a) also shows a scan at the equivalent wave vector $Q=(0.4, 0.2, 1)$ from the other chirality. The temperature dependence of the rocking curve scans demonstrates that the N$\acute{\mathrm{e}}$el temperature is approximately 425 K, which is significantly lower than that in the $A_{0.8}$Fe$_y$Se$_2$ system\cite{wbao1}. The fingerprint reflection peaks of the $\sqrt{5}\times\sqrt{5}$ iron vacancy order at $Q=(0.2, 0.6, 2)$ and $Q=(0.4, 0.8, 2)$ were also investigated and are represented in Fig.~\ref{fig3} (e, f). The order-disorder transition temperature of the iron vacancies occurs at 600 K. Here we have carried out $\theta$-$2\theta$ scans in order to track the temperature dependence of the iron vacancy order, while accounting for thermal expansion.

\begin{figure}[t]
\includegraphics[scale=0.6]{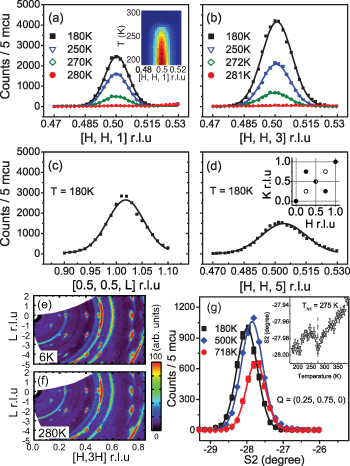}
\caption{(color online).  Diffraction studies on the in-plane AF phase with rhombic iron vacancy order. Scans of magnetic peaks at wave vector of (a) $Q=(0.5, 0.5, 1)$, (b) $Q=(0.5, 0.5, 3)$, (d) $Q=(0.5, 0.5, 5)$ along the $[H, H]$ direction, and (c) $Q=(0.5, 0.5, 1)$ along $L$ direction at selected temperatures. The inset of (a) is a color map of detailed temperature dependence. (e, f) show the nuclear peaks at the wave vectors of $H=0.25, 0.5, 0.75; L=even$ and magnetic peaks at $H=0.5, L=odd$ from the in-plane AF phase in the $[H, 3H, L]$ plane at 6 K and 280 K, respectively. The nuclear peaks were not changed, but the magnetic peaks disappeared at 280 K. The peaks at $H=0.4, L=even$ originate from the $\sqrt{5}\times\sqrt{5}$ iron vacancy order in block AF phase. (g) $\theta$-$2\theta$ scans at peaks of the rhombic iron vacancy order at $Q=(0.25, 0.75, 0)$ at 180 K, 500 K and 718 K. The evolution of the center of $S2$ versus temperature indicates an in-plane lattice constant change across the in-plane magnetic structure transition in the inset of (g).}
\label{fig4}
\end{figure}

Figure~\ref{fig4} summarizes the Bragg peaks of the `234' phase associated with the in-plane stripe AF order and rhombic iron vacancy order. The magnetic peaks are accurately centered at the wave vectors $Q=(0.5, 0.5, L=1, 3, 5)$ at 180 K with lattice constants $a=b=3.889$ \AA , $c=13.889$ \AA . The magnetic peaks disappear completely by 280 K. The rhombic iron vacancy order together with the stripe AF order will induce magnetic peaks at $Q=(0.25, 0.25, L=odd)$, $Q=(0.75, 0.75, L=odd)$; and nuclear peaks at $Q=(0.25, 0.75, L=even)$, $Q=(0.75, 0.25, L=even)$, $Q=(0.5, 0.5, L=even)$ as demonstrated in the inset of Fig.~\ref{fig4} (d). We show reflection peaks in the $[H, 3H, L]$ plane in Fig.~\ref{fig4} (e) at 6 K and (f) at 280 K. The peaks centered at $Q=(0.25, 0.75, L)$, $Q=(0.5, 0.5, L)$ and $Q=(0.75, 2.25, L), L=0, -2, -4$ are consistent with the rhombic iron vacancy order. The magnetic peak at $Q=(0.5, 1.5, 3)$ at 6 K in Fig.~\ref{fig4} (e) disappears at a temperature above $T_N=275$ K. The peaks at $Q=(0.4, 1.2, L=0, -2, -4)$ originate from the $\sqrt{5}\times\sqrt{5}$ iron vacancy order of the `245' phase. The temperature dependence of the $\theta$-$2\theta$ scans in Fig.~\ref{fig4} (g) shows the existence of the rhombic iron vacancy order at temperatures as high as 718 K; this is the reason that the two phases did not merge together at the temperature above the iron vacancy order-disorder transition at $T_S=600$ K of the `245' phase in Rb$_{0.8}$Fe$_{1.5}$S$_2$\cite{ricc1, scott}. From the inset, one can see a clear anomaly in the temperature dependence of the in-plane lattice constant at the AF transition indicating strong coupling between the structure and the antiferromagnetism. Residual peaks with temperature-independent intensities were observed at the magnetic peak positions above $T_N$ in semiconducting K$_{0.81}$Fe$_{1.58}$Se$_2$\cite{jzhao}. However, we did not observe residual intensity at these positions above $T_N$ in our semiconducting Rb$_{0.8}$Fe$_{1.5}$S$_2$ single crystals. This significant difference in these two systems, which otherwise behave quite similarly, remains to be understood.  

\begin{figure}[t]
\includegraphics[scale=0.4]{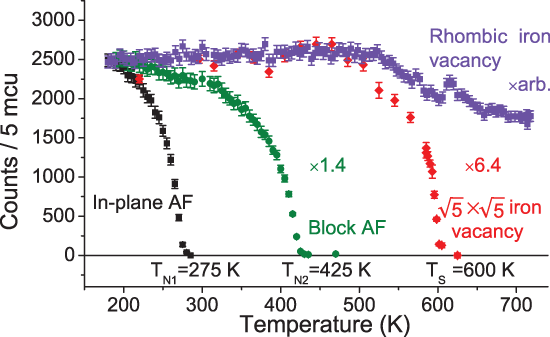}
\caption{(color online). Temperature dependence of order parameters for the in-plane stripe AF, the rhombic iron vacancy, the block AF and the $\sqrt{5}\times\sqrt{5}$ iron vacancy orders. The height of peaks at $Q=(0.5, 0.5, 1)$ fitted from $[H, H]$ scans shown as the black squares represent the stripe AF transition at $T_{N1}=275$ K. The temperature dependence of peak height at $Q=(0.2, 0.4, 1)$ (green circle) fitted from rocking curve scans shows the block AF transition at $T_{N2}=425$ K.  The red diamonds obtained from $\theta$-$2\theta$ scans through $Q=(0.4, 0.8, 2)$ indicate a first order-like transition of the $\sqrt{5}\times\sqrt{5}$ iron vacancies at   $T_{S}=600$ K. The rhombic iron vacancy order parameter integrated from the $\theta$-$2\theta$ scans at $Q=(0.25, 0.75, 0)$ was collected from another piece of single crystal with the same composition aligned in the $[H, 3H, L]$ zone.}
\label{fig5}
\end{figure}

In order to determine the transition temperatures of the `234' and `245' phases in Rb$_{0.8}$Fe$_{1.5}$S$_{2}$, we carefully measured the intensities of the fingerprint reflection peaks versus temperature; the results are shown in Fig.~\ref{fig5}. The N$\acute{\mathrm{e}}$el temperature of the in-plane stripe AF order of $T_{N1}=275$ K in the `234' phase of semiconducting Rb$_{0.8}$Fe$_{1.5}$S$_2$ is very close to $T_N=280$ K of the stripe AF order in semiconducting K$_{0.81}$Fe$_{1.58}$Se$_2$\cite{jzhao}. The block AF order of the `245' phase has a N$\acute{\mathrm{e}}$el temperature at $T_{N2}=425$ K and an iron vacancy ordering temperature of $T_{S}=600$ K. These have a much larger separation than those in the $A_{0.8}$Fe$_y$Se$_2$ system\cite{wbao1, fye}. 

\section{Discussions and Conclusions}

The similarity of the N$\acute{\mathrm{e}}$el temperatures suggests the crucial role of local moment superexchange interactions between the iron spins. Thus, we propose that strong correlation effects are essential to the formation of the stripe AF phase, in contrast with the spin-density-wave mechanism, which has been proposed as the origin of the magnetic order in the parent compounds of the iron pnictide superconductors\cite{jdong, mazin}. The strong coupling scenario can also be reconciled with the absence of hole Fermi surfaces in $A_{0.8}$Fe$_{y}$Se$_{2}$\cite{yzhang,xpwang,xjzhou}.  Similar to the iron pnictides, the spin resonance modes associated with superconductivity in iron chalcogenide (FeTe$_{1-x}$Se$_x$ and $A_{0.8}$Fe$_y$Se$_2$) systems are compatible with nesting between the hole-electron or electron-electron Fermi surfaces\cite{gfriemel}. In contrast with the pnictides, the in-plane magnetic orders in the iron chalcogenides are not compatible with nesting. The iron chalcogenides also have much larger local moments than the pnictide systems. The moments in the former are strongly suggestive of a localized rather than itinerant model for the magnetism. 

The data in Fig.~\ref{fig5} show a surprising feature which indicates that the two different structural phases are in communication with each other. Specifically, there is a small increase ($\sim$10$\%$) with increasing temperature in the intensity of the superlattice reflection associated with the rhombic vacancy order at the temperature at which the vacancies in the $\sqrt{5}\times\sqrt{5}$ phase become disordered. The intensity change suggests that the iron, partially occupied on the rhombic vacancy sites of the `234' phase below $T_S=600$ K, moves to the iron vacancy disordered `245' phase. The movement of iron vacancies between the two phases in Rb$_{0.8}$Fe$_{1.5}$S$_{2}$ suggests a possible way to understanding the complex relationship between the AF structures and the superconductivity in the $A_{0.8}$Fe$_y$Se$_2$ system. In the compounds with net composition $A_{0.8}$Fe$_y$Se$_2$ ($1.5<y<1.6$), the material stabilized is a combination of the semiconducting phase $A_{x}$Fe$_{1.5\pm\delta}$Se$_2$ (`234' phase) with in-plane stripe AF order and rhombic iron vacancy order together with the insulating phase, $A_{0.8}$Fe$_{1.6\pm\delta}$Se$_2$ (`245' phase) with the block AF order and $\sqrt{5}\times\sqrt{5}$ iron vacancy order. By adding more iron, only the volume fraction of the two phases is changed, that is, one traverses a first order two phase coexistence region between the `234' and `245' phases. This explains naturally why the N$\acute{\mathrm{e}}$el temperature of the in-plane stripe AF order is so stable. The block AF phase with $\sqrt{5}\times\sqrt{5}$ iron vacancy order, $A_{0.8}$Fe$_{1.6\pm\delta}$Se$_2$, (`245' phase) with $\delta=0$, represents an end point of the two-phase coexistence region. In this picture, by further increasing the iron content beyond $y=1.6$, the material then separates into a new iron rich superconducting, non-magnetic phase, and the block AF phase with $\sqrt{5}\times\sqrt{5}$ iron vacancy order. We speculate that the `245' phase is a stable stoichiometric phase and that the $\sqrt{5}\times\sqrt{5}$ ordered iron vacancies cannot be readily occupied. This means that increasing the Fe content above 1.6 causes the formation of a new iron-rich phase which exhibits superconductivity. Concomitantly, the iron-rich SC phase is always accompanied by the `245' phase but the `245' phase is not the parent compound of the superconducting phase. 

The results reported in this paper suggest a new strategy for probing the onset of superconductivity in the $A_x$Fe$_y$Se$_2$ type systems. In the pnictide systems, important insights have been gained by continuously tuning variables, such as the electron concentration by substitution (e.g., replacing Fe by Co or Ni) and thereby studying the evolution of the magnetism from the AF parent material to the superconducting material\cite{david}.  This is especially important at the onset of superconductivity where rich magnetic and superconducting behavior are observed.  This approach does not seem to be possible in the $A_x$Fe$_y$Se$_2$  systems since the superconductivity seems to appear discontinuously. Yet it is clear from the results reported here that systematic variation of the S content in the $A_{0.8}$Fe$_y$Se$_{2-z}$S$_z$ system should enable one to study the continuous evolution from the `parent' stripe AF sulphide to the superconducting mixed sulphide-selenide thus emulating studies in pnictide materials like those in BaFe$_2$As$_{2-x}$P$_x$\cite{matsuda}. The $A_{0.8}$Fe$_y$Se$_{2-z}$S$_z$ system may be closely analogous to the BaFe$_2$As$_{2-x}$P$_x$ system.

In summary, we have studied the magnetic and nuclear structures of semiconducting Rb$_{0.8}$Fe$_{1.5}$S$_{2}$ single crystals. Similar to semiconducting K$_{0.81}$Fe$_{1.58}$Se$_2$, there is an in-plane stripe AF phase with rhombic iron vacancy order, in addition to the block AF phase with $\sqrt{5}\times\sqrt{5}$ iron vacancy order. The robust $2.8\pm0.5 \mu_B$ in-plane ordered moments and $\sim$280 K N$\acute{\mathrm{e}}$el temperature of the stripe AF phase in semiconducting Fe-Se and Fe-S based systems suggest that strong electronic correlations play a dominant role in determining the nature of the magnetic state. The relationships among the block AF phase, the superconducting phase, and the in-plane stripe AF phase have been discussed in this paper. The $A_{0.8}$Fe$_{y}$Se$_{2-z}S_{z}$ system opens a new window to study the relationship between the antiferromagnetism and the superconductivity.

\section{Acknowledgements}   

This work is supported by the Director, Office of Science, Office of Basic Energy Sciences, U.S. Department of Energy, under Contract No. DE-AC02-05CH11231 and the Office of Basic Energy Sciences U.S. DOE Grant No. DE-AC03-76SF008. The research at Oak Ridge National Laboratory's High-Flux Isotope Reactor is sponsored by the Scientific User Facilities Division, Office of Basic Energy Sciences, U.S. Department of Energy. We also acknowledge support from the US NSF DMR-1362219 and the Robert A. Welch Foundation Grant No. C-1839 at Rice University.


\begin{thebibliography}{99}
\bibitem{discuss} The longitudinal scan at $Q=(1,1,0)$ on another piece of single crystal indicated $45\pm5\%$ `245' phase and $55\pm5\%$ `234' phase. The $(1,1,0)$ peaks from the two phases are separated as high as 718 K. Our neutron powder diffraction refinement on 2 grams sample ground by single crystals revealed that only 2$\%$ `245' phase and 98$\%$ `234' phase in the sample at 500 K.
\end{thebibliography}

\end{document}